\documentclass[manuscript,screen,nonacm]{acmart}
\AtBeginDocument{%
  \providecommand\BibTeX{{%
    \normalfont B\kern-0.5em{\scshape i\kern-0.25em b}\kern-0.8em\TeX}}}

\usepackage{multirow}
\usepackage{amsmath}
\usepackage{booktabs,caption,makecell,tabularx}
\usepackage{adjustbox}
\usepackage{caption}
\usepackage[flushleft]{threeparttable}
\begin{document}

\title{Topic-Centric Explanations for News Recommendation} 

\author{Dairui Liu}
\email{dairui.liu@ucdconnect.ie}
\orcid{0000-0002-8573-3857}
\affiliation{%
  \institution{Insight Centre for Data Analytics, School of Computer Science, University College Dublin}
  \city{Dublin}
  \country{Ireland}
}
\author{Derek Greene}
\email{derek.greene@ucd.ie}
\orcid{0000-0001-8065-5418}
\affiliation{%
  \institution{Insight Centre for Data Analytics, School of Computer Science, University College Dublin}
  \city{Dublin}
  \country{Ireland}
}
\author{Irene Li}
\email{ireneli@ds.itc.u-tokyo.ac.jp}
\orcid{0000-0002-1851-5390}
\affiliation{
  \institution{Information Technology Center, University of Tokyo}
  \city{Tokyo}
  \country{Japan}
}
\author{Xuefei Jiang}
\email{xuefei.jiang@ucdconnect.ie}
\orcid{0009-0007-3547-3114}
\affiliation{%
  \institution{Insight Centre for Data Analytics, School of Computer Science, University College Dublin}
  \city{Dublin}
  \country{Ireland}
}

\author{Ruihai Dong}
\email{ruihai.dong@ucd.ie}
\orcid{0000-0002-2509-1370}
\affiliation{%
  \institution{Insight Centre for Data Analytics, School of Computer Science, University College Dublin}
  \city{Dublin}
  \country{Ireland}
}


\begin{abstract}
News recommender systems (NRS) have been widely applied for online news websites to help users find relevant articles based on their interests. Recent methods have demonstrated considerable success in terms of recommendation performance. However, the lack of explanation for these recommendations can lead to mistrust among users and lack of acceptance of recommendations. To address this issue, we propose a new explainable news model to construct a topic-aware explainable recommendation approach that can both accurately identify relevant articles and explain why they have been recommended, using information from associated topics. Additionally, our model incorporates two coherence metrics applied to assess topic quality, providing measure of the interpretability of these explanations. The results of our experiments on the MIND dataset indicate that the proposed explainable NRS outperforms several other baseline systems, while it is also capable of producing interpretable topics compared to those generated by a classical LDA topic model. Furthermore, we present a case study through a real-world example showcasing the usefulness of our NRS for generating explanations.

\end{abstract}

\maketitle

\newcommand{\irene}[1]{\textcolor{red}{#1}}
\section{Introduction}

With the development of online news services, such as Google News, millions of users can acquire news information from convenient platforms, rather than directly from traditional media sources. However, it can be difficult for users to browse all available sources in order to find relevant articles which match their specific interests. This has motivated the development of personalized \emph{news recommender systems} (NRS), which aim to identify relevant news articles based on the personal interests of a given user. These recommendations can improve users' experience and save time when finding interesting news. This has led to considerable work~\citep{EmbeddingbasedNR2017Shumpei, DKN2018Wang, DAN2019QiannanZhu, LSTUR2019MingxiaoAn, NPA2019ChuhanWu, NRMS2019ChuhanWu, NAML2019ChuhanWu} focused specifically on improving the recommendation performance of these systems. However, the provision of \emph{explanations} for news recommendations has rarely been considered by researchers. This deficiency can lead to many problems~\cite{ExplainableRA2020Zhang}: 1) users may not trust results provided by the system for poor recommendations, since they are not aware of recommendation reasons; 2) the system may be less effective in persuading users to accept results; 3) this may further decrease the system's trustworthiness. Thus, providing explanation is critical in helping users to understand why corresponding news items have been presented to them by a NRS. Providing explanation can also be helpful for the system designer, allowing them to understand situations where the news recommendation process fails. The faithfulness of the generated explanations is particularly important in this context~\cite{jacovi2020towards,FaithfullyER2018Hoeve}. 
In summary, an effectively explainable NRS, which is generally based on a standard NRS, should accurately recommend news and explain those recommendations simultaneously. 
\begin{figure}
    \centering
    \includegraphics[width=\linewidth]{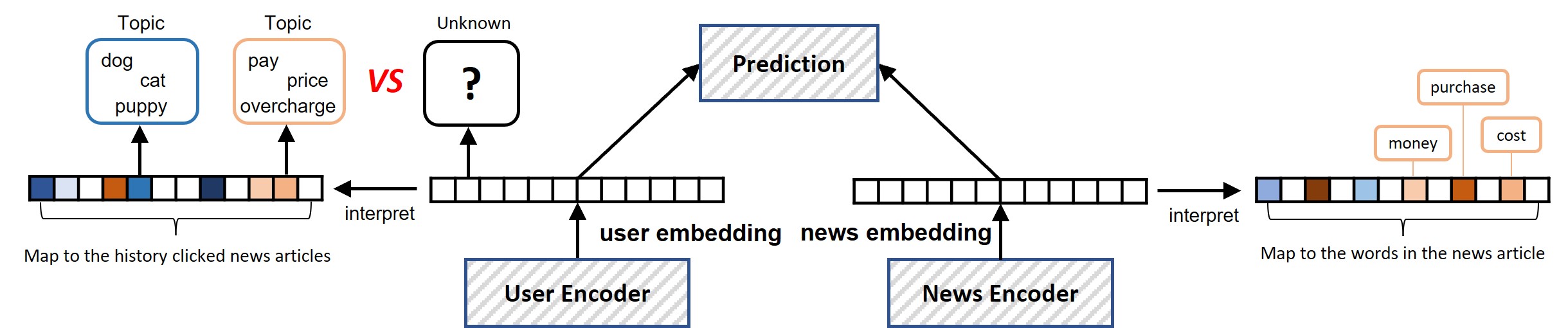}
    \caption{A simplified illustration of a standard recommender making a prediction based on the user embedding and the candidate news embedding from the respective encoders. Most existing NRSs only generate latent embeddings, making them difficult to understand. We aim to interpret these embeddings using latent topics. We map the latent feature of the user embedding to the corresponding history article and identify the topic descriptors of this article based on the corresponding news embedding.}
    \label{fig:recommender}
\end{figure}

The workflow of a standard personalized NRS involves several key steps~\cite{PersonalizedNRSurvey2021Wu}. The first step is to recall a small set of candidate articles from a large-scale news pool when a user visits the news platform. The recommender will then rank these candidate articles according to the user's interests, as encoded in their individual profile. Subsequently, the system will display the top-$k$ ranked articles to the user and records their behavior at last, which can be used for future recommendations. The NRS usually has no explicit user interest data (e.g. rating scores), so only implicit feedback (e.g. clicks) will be available to the system. Among these steps, the recommender is the core component of a standard NRS (see Fig.~\ref{fig:recommender}). This involves encoding the textual content of news and the corresponding user profile (e.g. history of clicked news) separately. News content modeling is important when attempting to build a high-performance NRS because clicked news articles usually reflect a user's interests. The predictions are generated through the collaborative contribution of user embedding and news embedding. However, the operation of the prediction process with a standard recommender is difficult to understand because the factors affecting the prediction are unclear. Thus, an intuitive approach to solve the problem is to identify the most critical factors determining whether the user will click on an article, which can lead to several benefits: 1) it can help the system designer to understand the underlying reasons behind the user's behavior, thereby allowing them to improve the recommendation process; 2) by presenting a transparent recommendation result to users, it can increase their trust and encourage them to accept the recommended news articles; 3) a more explainable system can lead to a better user experience, which helps build a trustworthy system.

\begin{figure}[!t]
    \centering
    \includegraphics[width=\linewidth]{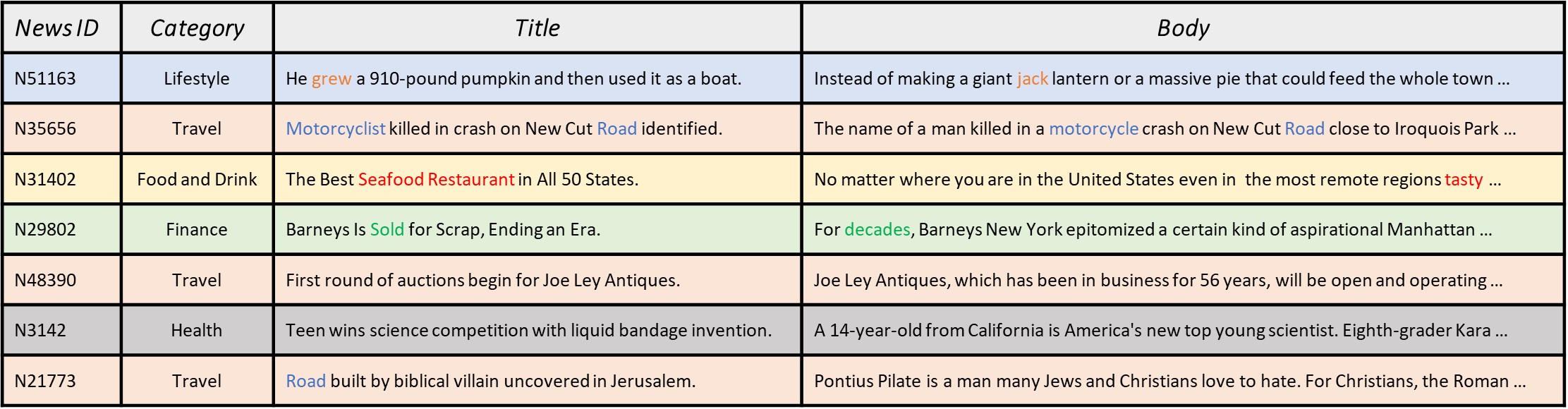}
    \caption{An example of the history sequence of news article clicks by a given user $U91836$, which reflects the broad interests of the user involving topics around travel, food, finance, and health. We highlight some topic indicators, such as "\textit{seafood}" and "\textit{restaurant}", for the food-related topic.}
    \label{fig:history_news_example}
\end{figure}

It is worth noting that news articles usually contain rich textual metadata, such as title, abstract, and body content, which can be leveraged as a valuable resource for providing explanations. In this work, we not only process news articles for recommendations, but also extract topics from them to interpret generated news embedding as shown in Fig~\ref{fig:recommender} to build an explainable NRS. In Fig~\ref{fig:history_news_example}, we see an example of the clicked news history from a random user from the real-world MIND dataset~\cite{wu-2020-mind}, which demonstrates the latent topics of the user's interests. The category is marked by the row color, while the topic indicators are highlighted by the font color. The selected user appears to have a broad range of interests, but is perhaps most interested in travel news. Thus, the user has a high probability of clicking on an article from the "travel" category. However, explaining recommendations using fixed category information might lack nuance and flexibility since a high-level news category will generally consist of many sub-topics which might evolve over time. Thus, discovering topics in the news corpus and employing them as explanations is a core objective of our explainable NRS. 

News content can be accurately encoded with the help of existing methods, such as the multi-head self-attention mechanism~\citep{NRMS2019ChuhanWu} or convolutional neural networks~\citep{LSTUR2019MingxiaoAn}. However, these methods are not explainable. Therefore we propose to embed an explainable news model~\citep{BATM2022Liu} that can extract explainable topics when encoding news content.
%
Specifically, we aim to provide explanations generated in a topic-centric manner, which differs from general explanations of recommendations~\citep{ExplainableRA2020Zhang}.
Compared to previous studies~\cite{FaithfullyER2018Hoeve, Liu2021ReinforcedAK, Zhang2022METoNRAM} on NRS explainability research, we generate explanations from a large news corpus with specific topic indicators, instead of simply providing a high-level category name. 
The key aspects of our work are as follows:
\begin{itemize}
    \item Most existing news recommendation system studies~\cite{NRMS2019ChuhanWu,NPA2019ChuhanWu,NAML2019ChuhanWu,LSTUR2019MingxiaoAn} focus on improving recommendation performance, while ignoring the interpretability of models and explanations about recommendations. Therefore, we propose to use an interpretable news encoding model--Bi-level Attention-based Topical Model(BATM)~\cite{BATM2022Liu} to learn an explainable news representation. Thus, we extract attention weights from multiple attention networks during the news modeling and user modeling procedure to generate topic-aware explanations. We achieve state-of-the-art performance on recommendation when compared with selected baselines, while also discovering interpretable topics from the news corpus.
    \item Although some researchers have presented visual explanations for recommendations to support model transparency, such work did not measure the interpretability of the model~\cite{NPA2019ChuhanWu,NAML2019ChuhanWu}. In other words, some case studies were provided to show interpretability, but missing quantitative evaluation metrics were not considered by the authors. Thus, we propose to use quantitative topic coherence metrics~\cite{NPMIVal2014MachineRT,word2vec2015OCallaghan} to evaluate the quality of topics extracted by our model, which reflects the interpretability of these explanations. Although our primary goal is around explanation rather than topic modeling, our experimental results show that we can often generate high-quality topics comparable to those generated by a standard LDA topic model~\cite{LDA2003blei}.
\end{itemize}

The remainder of this paper is organized into three parts: related work, methodology, and experimental results. Due to the lack of previous work around explainable NRS, we mainly concentrate on reviewing news recommendation methods in general in Section~\ref{sec:NR_methods}, with a short review of explainable methods in Section~\ref{sec:explainable_methods}. We compare the recommendation performance of our explainable NRS with several baselines in Section~\ref{sec:NR_performance}. In addition to evaluating recommendation performance, we use two topic coherence metrics to assess the extracted topics and compare them with a standard topic model in Section~\ref{sec:topic_eval}. Finally, in Section~\ref{sec:case_study} we present a case study showing the kinds of explanations generated by our approach for real recommendations. The source code associated with this work will be made available online\footnote{\url{https://github.com/Ruixinhua/ExplainedNRS}}.

\section{Related Work}

News recommendation, based on the personal interests of each user, is an active research area. Most researchers focus on the improvement of the performance of news recommendation systems, while few works consider recommendation explanations. In this section, we first introduce several popular approaches for news recommendation in Section~\ref{sec:NR_methods}. We then discuss explainable methods for the recommendation task involving text data in Section~\ref{sec:explainable_methods}.

\subsection{News Recommendation Methods} 
\label{sec:NR_methods}
Recommendation methods can be categorized into three broad branches:
\emph{collaborative filtering} (CF), \emph{content-based filtering} (CBF), and hybrid methods~\cite{GediminasAdomavicius2005TowardTN}. The CF approach makes predictions primarily based on the interaction between users and news, without knowing the news features in advance. However, this method often suffers from a severe cold start problem \cite{lam2008addressing}. CBF methods can address this problem by introducing the content associated with users and news, which is our primary research focus. Since it is often easier for researchers to solve a problem in news recommendations using CBF methods, they have become particularly popular in recent years. Many researchers have demonstrated that CBF is usually more effective than a pure CF method~\cite{ShainaRaza2021NewsRS}. Deep learning (DL) models have gradually become predominant in this area because of their performance when dealing with content-based news recommendation~\cite{ShainaRaza2021NewsRS}. Thus, with the success of DL-based CBF, we have looked at research works that share a similar technical paradigm~\cite{PersonalizedNRSurvey2021Wu}, including news modeling, user modeling, and news ranking procedures. Most recent studies~\cite{DKN2018Wang,LSTUR2019MingxiaoAn,NAML2019ChuhanWu,NPA2019ChuhanWu,NRMS2019ChuhanWu} have proven the effectiveness of the recommendation paradigm for modeling news and user representation separately. Next, we discuss successful methods for modeling news and user representation, some of which will provide baselines for our experiments.

\subsubsection{News Modeling} 
\label{sec:news_modeling}
The main goal of news modeling is to comprehend the characteristics and content of news, which is the core problem of the news recommender system. Due to the short lifespan of news items, the performance of CF methods~\cite{AbhinandanSDas2007GoogleNP} which only represent news articles by their IDs, is usually sub-optimal compared to CBF methods. Thus, we only focus on CBF news modeling methods, incorporating content features to represent news. These methods have traditionally extracted content-based features from the text of articles to construct a vector space model (VSM) representation \cite{salton1975vector}. Some approaches use handcrafted features, such as the concept frequency-inverse document frequency (CF-IDF)~\cite{Goossen2011NewsPU} and CF-IDF+~\cite{Koning2018NewsRW} models which extend standard TF-IDF term weighting schemes. However, these manually-designed methods usually require much effort and domain knowledge, which is also not optimal in understanding the semantic information among news texts.

To better encode the semantic meaning of news article content, dense embedding-based representations have been proposed as an alternative to sparse VSM models \cite{TomasMikolov2013DistributedRO}. Such modern models in natural language processing (NLP) can be helpful when encoding news content for recommendations. For instance, researchers have proposed an embedding-based news recommendation (EBNR) method~\cite{EmbeddingbasedNR2017Shumpei}, based on a variant of a de-noising autoencoder to learn representations from article texts. Other embedding-based methods, such as the deep structured semantic model (DSSM)~\cite{PoSenHuang2013LearningDS}, have applied a deep neural network (DNN) on existing embeddings to learn hidden news representations. However, most recent studies use the popular word2vec embedding method~\cite{TomasMikolov2013DistributedRO} for constructing dense vectors to represent the words appearing in news articles. These embedding-based models use simple DNNs to model news text, but can sometimes fail to capture contextual information accurately. Thus, some researchers have employed more complex neural networks, like a 3-D convolutional neural network (CNN)~\cite{VaibhavKumar2017WordSB} or a knowledge-aware CNN (KCNN)~\cite{DKN2018Wang}, to mine deep semantic relationship. For instance, DKN~\cite{DKN2018Wang} attempts to discover latent knowledge-level connections from news article titles by using a word-entity-aligned KCNN to learn a knowledge-enhanced news representation. This method also incorporates a knowledge graph (KG) to encode entities using KG embedding algorithms such as TransD~\cite{GuoliangJi2015KnowledgeGE}. Similarly, other methods also enhance news modeling by involving more CNNs, such as DAN~\cite{DAN2019QiannanZhu}, which adopts a combination of two parallel CNNs, built from news titles and named entities, respectively.

While these CNN-based methods can effectively learn contextual representations for news items, they are often not good at capturing and highlighting informative words because of the difference in news content informativeness. Thus, some authors have introduced attention mechanisms \cite{Bahdanau2015NeuralMT,vaswani2017attention} to address the problem. For example, NPA~\cite{NPA2019ChuhanWu} uses a personalized word-level attention-based CNN to learn attentive news representations. Similar to NPA, LSTUR~\cite{LSTUR2019MingxiaoAn} and NAML~\cite{NAML2019ChuhanWu} both learn a combined representation of multiple news-related metadata, including titles, categories, subcategories, and news body content through CNNs and attention networks. Here category and subcategory embeddings provide multi-view information to provide a better understanding of news content. Further studies employ advanced attention models like NRMS~\cite{NRMS2019ChuhanWu} and FedRec~\cite{TaoQi2020PrivacyPreservingNR} because of the effectiveness of attention mechanisms. NRMS~\cite{NRMS2019ChuhanWu} uses a multi-head self-attention (MHSA) network to capture word-level relations and applies another additive attention network to learn informative news representations. FedRec~\cite{TaoQi2020PrivacyPreservingNR} utilizes a news encoder with a combination of CNN, MHSA, and additive attention network to form a comprehensive representation of article titles. 

\subsubsection{User Modeling}
During the recommendation task, it is also essential to understand a user's interests from their profile, which usually consists of the history of their click behavior. Thus, user modeling is typically determined by modeling interactions, such as by inferring a user's interests from the sequence of their clicks. For example, EBNR~\cite{EmbeddingbasedNR2017Shumpei} considers a variant of the recurrent neural network (RNN)--gated recurrent unit (GRU) network, to generate user representations with history sequences. Similarly, RA-DSSM~\cite{VineetKumar2017DeepNA} adopts an attention-based bi-directional long short-term memory (Bi-LSTM, another variant of RNN) network to capture changing and diverse user interests. Moreover, LSTUR~\cite{LSTUR2019MingxiaoAn} applies a GRU network and ID embeddings for short-term and long-term user interest modeling respectively. 

In contrast, other methods like NAML~\cite{NAML2019ChuhanWu} and KRED~\cite{KRED2019DanyangLiu} only deal with the click sequences using a news-level attention network, which does not significantly affect the performance of recommendations. Similarly, DKN~\cite{DKN2018Wang} uses a candidate-aware attention network to form user representation by the relevance of clicked news and candidate news. And NPA~\cite{NPA2019ChuhanWu} considers using a personalized attention network with ID embeddings involved for user representation learning. Also, DAN~\cite{DAN2019QiannanZhu} employs the combination of attentive LSTM and candidate-aware attention network for user interest modeling. Moreover, NRMS~\cite{NRMS2019ChuhanWu} uses a more complex attention model, which is composed of a multi-head self-attention network and an additive attention network to learn contextual user representations. 

\subsection{Explainable Methods} 
\label{sec:explainable_methods}

Explainable recommenders aim to provide explanations for the recommendations they produce 
to indicate why a particular news item is being suggested to a user~\cite{ExplainableRA2020Zhang}. In recent years, with the success of deep learning, some models (i.e., NAML) have achieved impressive performance on news recommendations. However, these models are normally considered as a black box~\cite{EmbeddingbasedNR2017Shumpei, DAN2019QiannanZhu, DKN2018Wang, LSTUR2019MingxiaoAn, NPA2019ChuhanWu, NAML2019ChuhanWu, NRMS2019ChuhanWu}, making their outputs difficult to understand. Thus, we focus on explainable methods from the perspective of models themselves when making recommendations using text data.

One intuitive way to provide explanations is to look at attention scores. For example, the work by \citet{interpretable2017seo} applies a CNN to model review texts, and this method can show which part of the given review is more important for the output, based on attention scores. Similarly, some studies~\cite{contexta2019wu,coevlutionary2018lu} also consider the attention score over review words to explain recommendations. 
\citet{neurala2018chen} propose an approach to select relevant user reviews as explanations in their rating prediction model. They use an attention mechanism to analyze both user and item reviews, allowing them to identify high-quality reviews that can be used as explanations. Specifically, the attention module is applied to determine the usefulness of reviews, ensuring that only the most relevant and informative reviews are selected as explanations.
\citet{coevlutionary2018lu} integrate matrix factorization and an attentional GRU network on user-item rating data and customer reviews. The resulting user attention network is able to provide explanations by highlighting keywords and phrases based on attention scores. 
While the attention mechanism seems to be a straightforward method to help explainability, \citet{attention2019jain} argue that such a method may not be able to provide `meaningful' explanations.



In addition to the methods described above, researchers have also leveraged ideas from topic modeling to help improve explainable recommendations. The Latent Dirichlet Allocation (LDA) topic modeling~\cite{LDA2003blei} has been widely applied to investigate user preferences, which are often visualized using topic word clouds~\cite{hidden2013mcauley}. Other probabilistic graphic models are also studied for explainable recommendations. \citet{flame2015wuester} propose the Factorized Latent Aspect ModEl model (FLAME) that learns personalized preferences using item reviews. A word cloud is generated on the hotel aspect for hotel recommendations on the TripAdvisor corpus. Similarly, \citet{ZhaoCYZ15} utilize a probabilistic graphic model that leverages sentiment, aspect, and region information for point-of-interest (POI) recommendation, making it possible to provide personalized topical-aspect explanations. 

More recently, neural topic modeling \cite{DiengRB20} has been demonstrated to generate more useful topics when working with large, heavy-tailed vocabularies. \citet{tan2021panwar} propose the Topic Attention Networks for Neural Topic Modeling (TAN-NTM) framework. It first encodes a document using an LSTM module, and then a topic-aware module is applied to produce the outputs. A novel attention mechanism is used to learn topic-word distribution as well as the correlation of relevant words and the corresponding topic. This model shows promising results in both document classification and topic-guided keyphrase generation. To better align user preference and content information, \citet{Guo0LC22} propose the Topic-aware Disentangled Variational AutoEncoder (TopicVAE) model. This method first extracts topic-level item representations using an attention-based module and then adopts a variational autoencoder to model topic-level disentangled user representation. Experiments show that this model outperforms other selected baselines on recommendations, as well providing interpretability on disentangled representations. In this paper we also focus on generating explanations using a similar attention mechanism to extract topics for the news recommendation task.

\section{Methodology}  
\label{sec:methodology}
This section describes our proposed news modeling method~\cite{BATM2022Liu} and its application in the context of a general personalized news recommendations framework. We first formulate the recommendation problem and describe the process to generate explanations for the news recommendation scenario.  
\subsection{Problem Formulation}
\label{sec:problem_form}
Given a user $u$, let the browsing history (a set of news article documents) be denoted by $\mathcal{H}$, and the candidate news document set by $\mathcal{C}$. 
The history set consists of all their historical clicked news documents $\mathcal{H}= \{N_1, N_2, \ldots, N_i, \ldots, N_H\} $, where $H$ is the maximum recorded history length (i.e., we only keep the first $H$ news articles). The candidate set contains several news documents and their corresponding binary labels $\mathcal{C}=\{N_1\text{-}l_1, N_2\text{-}l_2, \ldots, N_i\text{-}l_i, \ldots, N_C\text{-}l_C\}$, where $C$ is the number of the candidate documents for the current impression ($C$ may be variant for different impressions) and $l_i$ reveals whether the user $u$ will click this news. We then calculate relevance scores $\mathcal{S}=\{s_1, s_2, \ldots, s_i, \ldots, s_C\}$ for $u$ and recommend the top-ranked news based on these scores. In this setting, the first problem (RQ1) is how to acquire an accurate ranked list of candidate documents to match the click preference of $u$. 
The second associated problem (RQ2) is how best to explain the ranking based on the information available. 
\begin{figure}[h]
  \centering
  \includegraphics[width=\linewidth]{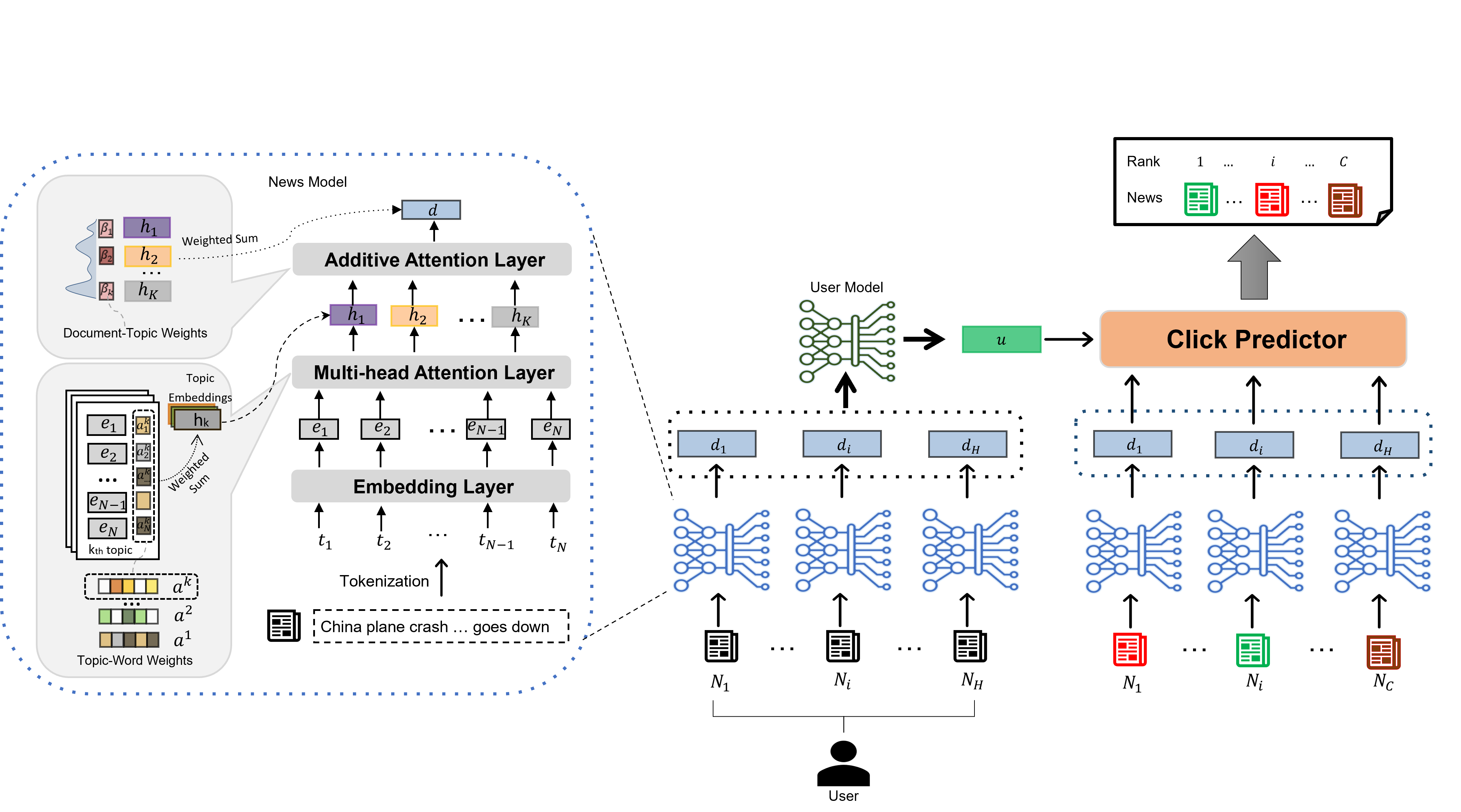}
  \caption{The architecture of the proposed news recommendation framework, which consists of a model that encodes both news articles and users.}
  \label{fig:architecture}
  \Description{News Recommendations Framework}
\end{figure}
\subsection{News Recommendations}
We now introduce the proposed news recommendation workflow and its key components, as illustrated in Fig~\ref{fig:architecture}. We first describe the encoder used for news modeling and then take the output of the news encoder to model the user representation. At the same time, the news encoder also generates candidate news representations, which provide the input for the ranking module, along with the user representation. Finally, we outline the training strategy and loss function used to train the recommendation system.

\subsubsection{News Modeling}
\label{sec:news_encoder}
This task aims to learn semantic news representations from documents using a shared news encoder module. Our proposed encoder, the Bi-level Attention-based Topical Model (BATM)~\cite{BATM2022Liu}, contains an embedding layer and two attention layers, which can generate news embeddings for recommendations and extract meaningful topics for explanations. For a given input document $N_i$, we first tokenize the document into a set of tokens $T_i=\{t_1, t_2, \ldots, t_n, \ldots, t_N\}$, where $N$ is the maximum length of the tokenized set. Then we convert tokens $T_i$ to embedding vectors $E_i=\{e_1, e_2, \ldots, e_n, \ldots, e_N\}$ in the embedding layer, where the embedding $e_n$ represents word vector of the token $t_n$. We use pre-trained word embedding vectors to enhance the semantic meanings accordingly, improving recommendations' performance and topics' quality.

After the embedding layer, we pass word vectors $E_i$ into two attention layers. The first one utilizes a multiple-topic attention mechanism to allow the model to focus on different positions in the document from different representation subspaces. We employ $K$ attention heads to capture $K$ topics among news corpus by topic-term weights $\mathcal{A}$, and here is the procedure of this layer: 1). we compute attention values ${g}^{k}$ through feed-forward neural networks as shown in Eq.~\ref{eq:1}; 2) then use the \textit{softmax} function to get the distribution of the normalized weights; 3) finally, we calculate the weighted sum of word embedding vectors using the attention weights to acquire the topic vector $h_{k}$:
\begin{equation}
\label{eq:1}
\mathrm{g}_{j}^{k}=v_{k}^{T} \tanh \left(\mathrm{W}_{K} e_{j}+ b_{k}\right)
\end{equation}
\begin{equation}
\label{eq:2}
\alpha_{j}^{k}=\frac{\exp({\mathrm{g}_{j}^{k}})}{\sum_n^N \exp({\mathrm{g}_n^{k}})}
\end{equation}
\begin{equation}
\label{eq:3}
h_k=\sum_{j}^N \alpha_{j}^{k} e_{j} 
\end{equation}
In the above the learnable parameters are $
v_{k} \in \mathbb{R}^{D_{K}}$, $\mathrm{~W}_{K} \in \mathbb{R}^{D_{E} \times D_{K}}$, and ${b_k} \in \mathbb{R}^{D_{K}}$, where $D_K$ is the projected dimension of each attention in the middle and $D_{E}$ is the embedding dimension of word vectors. We extract normalized attention weights $\mathcal{A}=\{\alpha^1, \alpha^2, \ldots, \alpha^k, \ldots, \alpha^K\}$ as topic-term weight distribution, where $\alpha^k \in \mathbb{R}^N$. The output topic vectors are $\{h_1, h_2, \ldots, h_k, \ldots, h_K\}$, where $h_k \in \mathbb{R}^{D_H}$ and ${D_H}$ is the dimension of the topic vectors and document representations.

Finally, we feed the topic vectors into the additive attention layer, which generates the document-topic distribution $\mathcal{B}$ and the document representation $d_i$. This is achieved as follows: 1) we compute document topic attention values $\mu_{k}$ by Eq.~\ref{eq:4}; 2) we normalize attention weights by \textit{softmax} function to get document-topic weights; 3) we acquire the document vector $d_i$ through weighted sum up topic vectors according to normalized weights:
\begin{equation}
\label{eq:4}
\mu_{k}=V_I^T \tanh \left(\mathrm{W}_{I} h_{k}+b_{I}\right)
\end{equation}
\begin{equation}
\label{eq:5}
\beta_{k}=\frac{\exp({\mu_{k}})}{\sum_{k}^{K} \exp({\mu_{k}})} 
\end{equation}
\begin{equation}
\label{eq:6}
d_i=\sum_{k}^{K} \beta_{k} h_{k}  
\end{equation}
The trained parameters here are  $V_I \in \mathbb{R}^{D_I}, \mathrm{W}_{I} \in \mathbb{R}^{D_E \times D_I}$, and $b_I \in \mathbb{R}^{D_I}$, where $D_I$ is projected dimension of additive attention. Document-topic weights $\mathcal{B}^i=\{\beta_1, \beta_2, \ldots, \beta_k, \ldots, \beta_K\}$ reflects importance of each topic to the specific news $N_i$. The news representation $d_i \in \mathbb{R}^{D_H}$ is the final output of the news encoder, where the users' history news representations are used to form the user representation. 

\subsubsection{User Modeling}  
\label{sec:user_modeling}
This aspect is another core component of learning a user representation from the browsing history news $\mathcal{H}=\{d_1, d_2, \ldots, d_i, \ldots, d_H\}$. Inspired by previous studies~\cite{NAML2019ChuhanWu,NPA2019ChuhanWu,NRMS2019ChuhanWu}, we use an additive attention network to encode news history. The procedure for acquiring a user representation is similar to Eqs.\ref{eq:4}~to~\ref{eq:6}. Specifically, we calculate user-news attention weights $\gamma$ by normalizing attention values $\theta_{i}$ with a \textit{softmax} function. Then the user vector $u$ is calculated as the weighted sum of the user's history news representations using user-news attention weights. 
\begin{equation}
\label{eq:7}
\theta_{i}=V_U^T \tanh \left(\mathrm{W}_{U} d_i+b_{U}\right)
\end{equation}
\begin{equation}
\label{eq:8}
\gamma_{i}=\frac{\exp({\theta_{i}})}{\sum_{j}^{H} \exp({\theta_{j}})} 
\end{equation}
\begin{equation}
\label{eq:9}
u=\sum_i^H \gamma_{i} d_{i}  
\end{equation}
Here $V_U^T \in \mathbb{R}^{D_U}, {W}_{U} \in \mathbb{R}^{D_H \times D_U}$, and $b_U \in \mathbb{R}^{D_U}$ are trainable parameters, and $D_U$ is the dimension of the projection layer. We determine the significance of the news item $N_i$ via the user-news weight $\gamma_{i}$, which is also used to select the most relevant news articles for the recommendations task.

In addition to attentive methods, sequential methods, such as Gated Recurrent Unit (GRU) networks~\cite{GRU2014Cho}, can also be effective in modeling a user's interest~\cite{EmbeddingbasedNR2017Shumpei,LSTUR2019MingxiaoAn}. Thus, we further consider the impact of using a GRU network as the user model, where the user vector is computed as follow~\cite{GRUDoc}:
\begin{flalign}
\label{eq:10-13}
&r_t=\sigma(W_{ir} d_t + W_{hr} o_{t-1} + b_r) \\
&z_t=\sigma(W_{iz} d_t + W_{hz} o_{t-1} + b_z) \\
&n_t=\tanh \left(W_{in} d_t + b_{in} + r_t*(W_{hn} o_{t-1} + b_{hn})\right) \\
&o_t=(1-z_t)*n_t+z_t*o_{t-1}
\end{flalign}
Accordingly, $W_{ir}, W_{hr}, W_{iz}, W_{hz}, W_{in}, W_{hn} \in \mathbb{R}^{D_H \times D_H}$ and $b_r, b_z, b_{in}, b_{hn} \in \mathbb{R}^{D_H}$ are learnable hidden weights and biases. We represent the input news representation at time $t$ with $d_t$, the hidden states at time $t-1$ and $t$ with $o_{t-1}$ and $o_t$, respectively, and the reset, update, and new gates at time $t$ with $r_t$, $z_t$, and $n_t$, respectively. $\sigma$ and $*$ are the sigmoid function and the Hadamard product.

We denote the first strategy that uses additive methods as the $BATM\text{-}ATT$ model, and the second strategy incorporating the GRU network as the $BATM\text{-}GRU$ model. The output user representation $u \in \mathbb{R}^{D_H}$ reflects a user's interest, which will be employed to determine the click probabilities of corresponding groups of candidate news articles. 

\subsubsection{Click Predictor and Training Strategy}  
By obtaining the user representation $u$ and the set of candidate news representations $D=\{d_1, d_2, \ldots, d_i, \ldots, d_C\}$, we can employ the simple inner product to calculate the news click probability score, which is inspired by recent research~\cite{NRMS2019ChuhanWu,NPA2019ChuhanWu,NAML2019ChuhanWu,LSTUR2019MingxiaoAn}. The probability score $s_i$ of the news $N_i$ is computed as $s_i=u^Td_i$, which determines the recommendation rank among the candidate news items. For the model training procedure, we use the negative sampling strategy~\cite{NRMS2019ChuhanWu,PoSenHuang2013LearningDS} to train our ranking model using the NCE loss. We treat each clicked news item as a positive sample of the candidate news set, and randomly select $M$ non-clicked news items from the same impression as negative samples. Then we jointly calculate scores of positive and negative to acquire the NCE loss. Finally, the loss $\mathcal{L}_{NCE}$ is given by the negative log-likelihood of all positive samples of this impression:
\begin{equation}
\label{eq:14}
\mathcal{L}_{NCE}=-\sum_p^P\log {\frac{\exp(s_p^+)}{\exp(s_p^+)+\sum_m^M \exp(s_m^-)}}
\end{equation}
where $P$ represents the number of positive training samples in the impression and $s_m^-$ denotes the $m^\text{th}$ negative sample in the same session linked to the $p^\text{th}$ positive sample.

\subsection{Recommendation Explanations}
\label{sec:rec_explain}
After training the proposed recommendation system, we analyze the attention weights extracted from the trained $BATM\text{-}ATT$ only to generate explanations because the $BATM\text{-}GRU$ model is not fully explainable. An explanation is provided to clarify why the model recommends such news items, and we usually only care about the several foremost recommended news items. The core aspect of our explanations are the extracted topics used to reveal the relatedness between the user's browsing history news and the ranked candidate news. Thus, we propose to use the quantitative topic coherence metrics~\cite{NPMIVal2014MachineRT,word2vec2015OCallaghan} to globally evaluate the quality of topics extracted by the $BATM\text{-}ATT$ model in Section~\ref{sec:topic_eval}, which can reflect the trustworthy of these explanations. Later in Section~\ref{sec:case_study} we present a case study to further motivate the idea of using topics to validate the relevance of related news articles in a real-world example.
\begin{figure}[!ht]
  \centering
  \includegraphics[width=\linewidth]{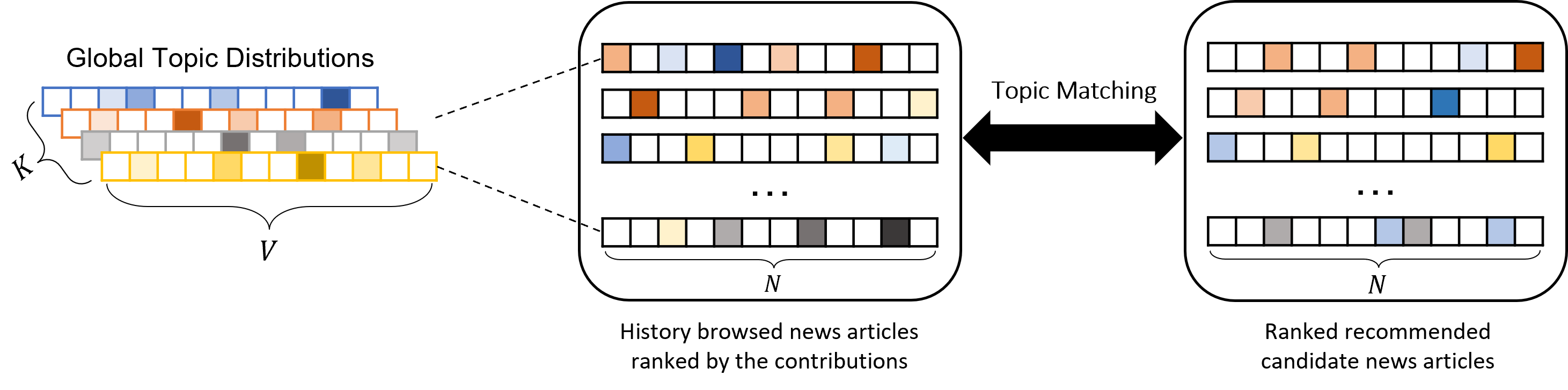}
  \caption{An illustration of the process of using topics as recommendation explanations, where different colors represent different topics. The more saturated color indicates a higher topic weight for a given word.}
  \label{fig:explanations}
  \Description{News Recommendations Explanations}
\end{figure}

The generation of explanations consists of two steps, as shown in Fig.~\ref{fig:explanations}. Here we revisit the recommendation example previously discussed in Section~\ref{sec:problem_form}. The first phase involves the global topic distribution from the multiple-topic attention layer, as calculated by Eq.~\ref{eq:1} and Eq.~\ref{eq:2} using embedded word vectors as inputs. Assume that there are $K$ global topics in total and $V$ words in the corpus, so the resulting topic distribution $\mathcal{T}$ is a $K \times V$ weight matrix. Moreover, we extract the $top\text{-}M$ most important words from the global topic distribution $\mathcal{T}$ for each topic and view them as being the topic's \emph{descriptors}, which are used for the topic's quality evaluation. The next phase involves identifying the news contribution among the browsing history $\mathcal{H}$ for the candidate set $\mathcal{C}$. Here we can recognize the news contribution by using user-news attention weights $\gamma$ as calculated by Eq.~\ref{eq:7} and Eq.~\ref{eq:8}. A news article from $\mathcal{H}$ should correlate highly with the clicked candidate news articles when they focus on similar topics. We select the most relevant topics according to the document-topic distributions $\mathcal{B}$ for each news article. To verify the topic relevance among the most contributed history clicked news articles and recommended news articles, we highlight the highest-scoring words from a subset topic distribution $\mathcal{T}$, which takes the form of a $K \times N$ matrix where $N$ is the length of the corresponding news article. Thus, we can validate the relevance of the history $\mathcal{H}$ and candidate $\mathcal{C}$ from the semantic meaning of those highlighted words.

\section{Experiments}

We now provide details of experimental evaluations, including a description of the news dataset, baseline models, and experimental settings for the news recommendation task. We also provide an evaluation of the global topics generated by our approach using topic coherence metrics, together with a case study in Section~\ref{sec:case_study}.
\subsection{Data} 
\label{sec:dataset_description}
We evaluated our proposed model on a news classification task~\cite{BATM2022Liu} and a news recommendation task on a real-world news recommendation dataset, MIcrosoft News Dataset (MIND)~\cite{wu-2020-mind}. MIND is a large-scale English news recommendation collection that consists of 1 million anonymized users and more than 160k English news articles, retrieved during 6 weeks from October 12 to November 22, 2019. In addition to the articles themselves, over 15 million impression logs were collected during this time period, involving more than 24 million user clicks. Each impression log represents a one-time recommendation that includes the IDs of news shown to a user when the user browses the news platform during a specific time slot, along with the click behaviors on these news articles. The content related to each news article includes its title, abstract, category, and URL. Based on the original news dataset, we add further information by including news body content for our experiments.
\begin{table}[!ht] 
    \begin{threeparttable}
    \caption{Summary information of the versions of MIND considered in our experiments.}
    \begin{tabular}{cccccccc}
        \toprule
        \textbf{MIND Version} &\textbf{\#Users} & \textbf{\#Impressions} & \textbf{\#Clicks} & \textbf{\#News} & \textbf{\#Avg. Len}  \\ 
        \midrule
        MIND-OFFICIAL & 1,000,000 & 15,777,377 & 24,155,470 & 161,013 & 639.57 \\  
        MIND-LARGE & 750,434 & 2,609,219 &  3,958,501 & 130,379 &         593.56 \\  
        MIND-SMALL & 94,057 & 230,117 &   347,727 &  65,238 & 638.43 \\  
        \bottomrule
    \end{tabular}
    \begin{tablenotes}
        \small 
        \item \#Avg. Len is the average length of the news content, including the title, abstract, and body. MIND-Official statistic data is from the MIND~\cite{wu-2020-mind}. MIND-LARGE and MIND-SMALL are publicly available versions. 
    \end{tablenotes}
    \label{tab: MIND_stat}
    \end{threeparttable}
\end{table}

The final publicly-released MIND dataset only contains two weeks' impression logs from November 9 to November 22, 2019, where only the first week's logs are labeled. MIND released two versions of the dataset from the first week's data, named MIND-LARGE and MIND-SMALL respectively. The larger dataset contains all behaviors from the first week, while the smaller one only includes the impression logs for a subset of the users. We list details of the official MIND version and publicly available version in Table~\ref{tab: MIND_stat}. 

In our evaluations, we take the MIND-LARGE dataset for the news classification (NC) task and randomly divide the data into training/validation/test sets. As for the news recommendation (NR) task, we randomly split the first week's log into three sets for both MIND-SMALL and MIND-LARGE versions: training set (from November 9 to November 14), validation set (November 15), and test set (November 15). We select the best hyperparameters for different models based on the validation set, and compare their relative performance on the test set. 

\subsection{Baseline Models and Experimental Settings}
\label{sec:baselines_setting}
For the purpose of assessing recommendations performance, we compare our model with several popular deep neural models designed for news recommendation:
\begin{itemize}
    \item DKN~\cite{DKN2018Wang} applies a word-entity-aligned KCNN on news representations learning and a candidate-aware attention network for recommendations\footnote{We use dot product instead of a linear neural network to predict click probability for a fair comparison.}.
    \item NAML~\cite{NAML2019ChuhanWu} leverages two separate CNN to
    encode news title and body text while using linear layers to encode category and subcategory. The news representations are learned from a multi-view of text, category, and subcategory representations. And modeling user behavior using another attention network.  
    \item NRMS~\cite{NRMS2019ChuhanWu} uses multi-head self-attentions for both news and user modeling.
    \item LSTUR~\cite{LSTUR2019MingxiaoAn} involves an attention-based CNN on news representations learning and a GRU network to model short-term user interest along with user id for long-term user interest.
    \item NPA~\cite{NPA2019ChuhanWu} uses a personalized word-level attention-based CNN to learn news representations, while another personalized attention network is employed to learn user representations.
\end{itemize}
We divide experimental configurations into two categories: general system settings and model hyperparameter settings. Consistent with previous studies~\cite{DKN2018Wang,NPA2019ChuhanWu,NAML2019ChuhanWu,LSTUR2019MingxiaoAn,NRMS2019ChuhanWu}, we maintain the same system settings across all experiments to ensure fairness. To account for GPU memory limitations, we set the maximum number of clicked news items for user representation learning to 50 and the maximum length of a news article to 100 tokens, with a maximum length of 30 for the news title and 70 for the news body, respectively. Thus, we randomly sampled at most 50 browsed articles from a user's history. The batch size was set to 64 and the negative sampling ratio $M$ was set to 4 during training, so that each positive sample is paired with 4 negative samples. We employed Adam~\cite{Kingma2014AdamAM} as the model optimization technique during gradient descent. The learning rate was set to 1e-3, and we reduced the learning rate by half each epoch. To mitigate overfitting, we added a dropout layer~\cite{Srivastava2014DropoutAS} to each layer for all models in comparison and set the dropout rate to 20\%. 
We initialized the embedding layer with pre-trained Glove embedding~\cite{glove2014Pennington} and set the embedding dimension $D_E$ to 300. Following previous work~\cite{wu-2020-mind}, we consider four ranking metrics in our experiments to evaluate the performance of news recommendations: AUC, MRR, nDCG@5, and nDCG@10. We repeat each experiment independently 5 times and select the model hyperparameters with the highest average AUC score on the validation set across those 5 runs. The performance scores on the test set are reported in Section ~\ref{sec:NR_performance}.

\subsection{News Recommendations Performance}
\label{sec:NR_performance}
The overall performance of all baselines and two variants of our model are summarized in Table~\ref{tab: performance}. All the numbers in the table are percentage numbers with ‘\%’ omitted. The overall best and previous best results are boldfaced and underlined respectively.

\begin{table*}[!ht] 
  \caption{Recommendation performance for different models, in terms of AUC, MRR, nDCG@5, and nDCG@10.}
  \label{tab: performance}
  \begin{adjustbox}{width={\textwidth},totalheight={\textheight},keepaspectratio}
  \begin{tabular}{@{} l|cccc|cccc @{}}
    \hline
    \multirow{2}{*}{\textbf{Models}} & \multicolumn{4}{|c}{\textbf{MIND-SMALL}} & \multicolumn{4}{|c}{\textbf{MIND-LARGE}} \\
    \cline{2-9} 
    & AUC & MRR & nDCG@5 & nDCG@10 & AUC & MRR & nDCG@5 & nDCG@10  \\
    \hline
    LSTUR~\cite{LSTUR2019MingxiaoAn} &    \underline{67.26±0.13} &    31.44±0.16  &   \underline{35.20±0.18}  &   41.43±0.19 &      \underline{69.31±0.16} &     33.38±0.22 &   37.39±0.21 &    43.57±0.17 \\
    NAML~\cite{NAML2019ChuhanWu} &    67.14±0.20  &    \underline{31.58±0.28}  &   35.20±0.29  &   \underline{41.52±0.28} &      69.24±0.17 &     \underline{33.96±0.27} &   \underline{37.85±0.21} &    \underline{44.02±0.18} \\
    NPA~\cite{NPA2019ChuhanWu} &    66.24±0.25 &    31.06±0.13  &   34.37±0.19 &   40.69±0.15 &      68.95±0.21 &     33.37±0.37 &   37.26±0.35 &    43.39±0.34 \\
    NRMS~\cite{NRMS2019ChuhanWu} &    66.58±0.17 &    31.44±0.15  &   34.99±0.19 &   41.21±0.16 &      69.09±0.13 &     33.25±0.42 &   37.19±0.33 &    43.43±0.29 \\
    DKN~\cite{DKN2018Wang} &    66.95±0.25 &    31.12±0.28  &   34.94±0.29 &   41.13±0.29 &      68.89±0.11 &     33.24±0.15 &   37.26±0.11 &    43.43±0.11 \\
    \hline
    BATM-ATT  &  \textbf{67.79±0.22} &    \textbf{32.44±0.28} &    \textbf{36.10±0.30} &    \textbf{42.29±0.26} &       \textbf{69.67±0.14} &     33.76±0.33 &    37.62±0.03 &    43.93±0.26 \\
    BATM-GRU & 67.48±0.22 &    32.12±0.39 &  35.76±0.44 &   41.95±0.36 &        69.63±0.10 &      \textbf{33.96±0.16} &    37.82±0.15 &     \textbf{44.06±0.15} \\
    \hline
\end{tabular}
\end{adjustbox}
\end{table*}

From the results, we can make a number of important observations.
First, all deep neural models achieved similar performances based on the experimental setting in Section~\ref{sec:baselines_setting}. This is because we used the same pre-trained embedding parameters (Glove) to initialize the embedding layer, which means the number of trainable parameters of models is close\footnote{The number of embedding layer's parameters usually counts 90\% of the model in total}. Another reason is that we applied a similar architecture which contains a news encoder and a user encoder and makes predictions using the results of dot-product between news representations and user representations. Thus, the difference in results is due to news modeling and user modeling design. 

Secondly, the performances of all models in the MIND-LARGE set are significantly better compared to the results in the MIND-SMALL set. NAML model generally achieved the best results among the selected baselines, except in some cases (e.g., the AUC metric) where LSTUR model performs slightly better than the NAML model, as it can obtain information from news text, category, and subcategory to form informative representations. LSTUR model is also competitive, as it uses a sequence model (GRU) for user modeling, effectively capturing user interests, which is also the reason we tried a GRU variant of our model.

Finally, our models $BATM\text{-}ATT$ and $BATM\text{-}GRU$ outperform all baselines on the MIND-SMALL set and the MIND-LARGE set for almost all metrics. We can observe that NAML model presents a comparable performance compared to our models, though our models are slightly better overall. The encoding of category and subcategory in NAML model may fuse explicit topic information into the final news representations, which is very similar to our topic modeling module. However, our model can extract latent topics from the news texts and acquire topic representation from texts instead of categories and subcategories. Overall, these experiments demonstrate the effectiveness of our proposed models for news recommendation.

\subsection{Evaluation of Global Topics}
\label{sec:topic_eval} 
In addition to performance on the news classification (NC) and news recommendation (NR) tasks, we also evaluate the usefulness of the explanations generated by our methods. As our explanations are based on extracted topics, we quantitatively assess usefulness here by considering the popular idea of \emph{topic coherence}, corresponding to the semantic interpretability of the top terms typically used to describe the topics identified by a topic modeling algorithm~\cite{word2vec2015OCallaghan}. We use this to compare  the quality of topics from our trained model with those produced by a classical topic model LDA~\citep{LDA2003blei}. Specifically, the trained model here is $BATM\text{-}ATT$ model, where the news corpus $\mathcal{S}_{large}$ for the training of $BATM$ model~\cite{BATM2022Liu} on the NC task. The LDA model is generated on the MIND-LARGE corpus, as described in Section~\ref{sec:dataset_description}. We also train the $BATM\text{-}ATT$ for NR task on the MIND-SMALL dataset and compare it with the model on NC task and the LDA model regarding topic quality. The news corpus $\mathcal{S}_{large}$ is also used as the reference corpus for the remaining assessments when calculating topic coherence scores.

After training the model, we calculate the topic coherence scores using two different metrics from the topic modeling literature: $NPMI$~\cite{NPMI2009Bouma,NPMIVal2014MachineRT} and $W2V$~\cite{word2vec2015OCallaghan}. These are used to determine whether the extracted topic descriptors have an intuitive meaning. The reason for using these two metrics is that the topic coherence metric $NPMI$ is regarded as positively correlated with human intuition~\citep{NPMIVal2014MachineRT}, while $W2V$ similarity is designed for embedding-based methods. We select the $top\text{-}M$ highest scoring words as topic descriptors and calculate $NPMI$ scores and $Word2Vec$($W2V$) similarity scores of them. For a given topic $k$, suppose we obtain a topic descriptor set $T_k=\{t_1^k, t_2^k, \ldots, t_n^k, \ldots, t_{N}^k\}$, so we compute the $NPMI$ scores and $W2V$ similarity scores as follow:
\begin{equation}
    NPMI(T_k)=\frac{1}{\binom M2} \sum_{j=2}^N \sum_{i=1}^{j-1} \frac{\log \frac{P\left(t_j^k, t_i^k\right)+\epsilon}{P\left(t_i^k\right) P\left(t_j^k\right)}}{-\log P\left(t_i^k, t_j^k\right)+\epsilon}
\end{equation}

\begin{equation}
    \label{eq:w2v}
    W2V(T_k)=\frac{1}{\binom M2} \sum_{j=2}^N \sum_{i=1}^{j-1} similarity(e_j^k, e_i^k)
\end{equation}
Here $e_j^k$ and $e_i^k$ in Eq.~\ref{eq:w2v} are the word vectors of tokens $t_j^k$ and $t_i^k$ from the trained model. We set $M=10$ for each topic and take the average values of all topics. The word co-occurrence probabilities of $t_j^k$ and $t_i^k$ are counted from the reference corpus $\mathcal{S}_{large}$. 

Corpus preprocessing can influence the resulting downstream topic descriptors and can further affect  topic evaluation metrics. The processing pipeline of a topic model mainly involves three steps: 1) filter out stopwords using the default spaCy English stopword list\footnote{\url{https://github.com/explosion/spaCy/blob/v3.0.5/spacy/lang/en/stop\_words.py}}; 2) remove tokens that appear in more than 90\% of documents; 3) remove tokens that appear in fewer than $N$ documents. We use the pipeline as preprocessing procedure to train an LDA model and evaluate it with the same processing pipeline. However, the preprocessing pipeline of our deep models is quite different as we keep most tokens (e.g., stopwords remained) for training to retain semantic information. Thus, we apply the same pipeline as the LDA model after the training process of NC and NR tasks. As it is performed after training,  we name it post-processing (denoted as $PP\text{-}N$, such as $PP\text{-}10$) for our $BATM\text{-}ATT$ model. For the model evaluated on the original training dataset, we denote it as $Without\text{ }PP$.

\begin{figure}[!t]
  \centering
  \includegraphics[width=\linewidth]{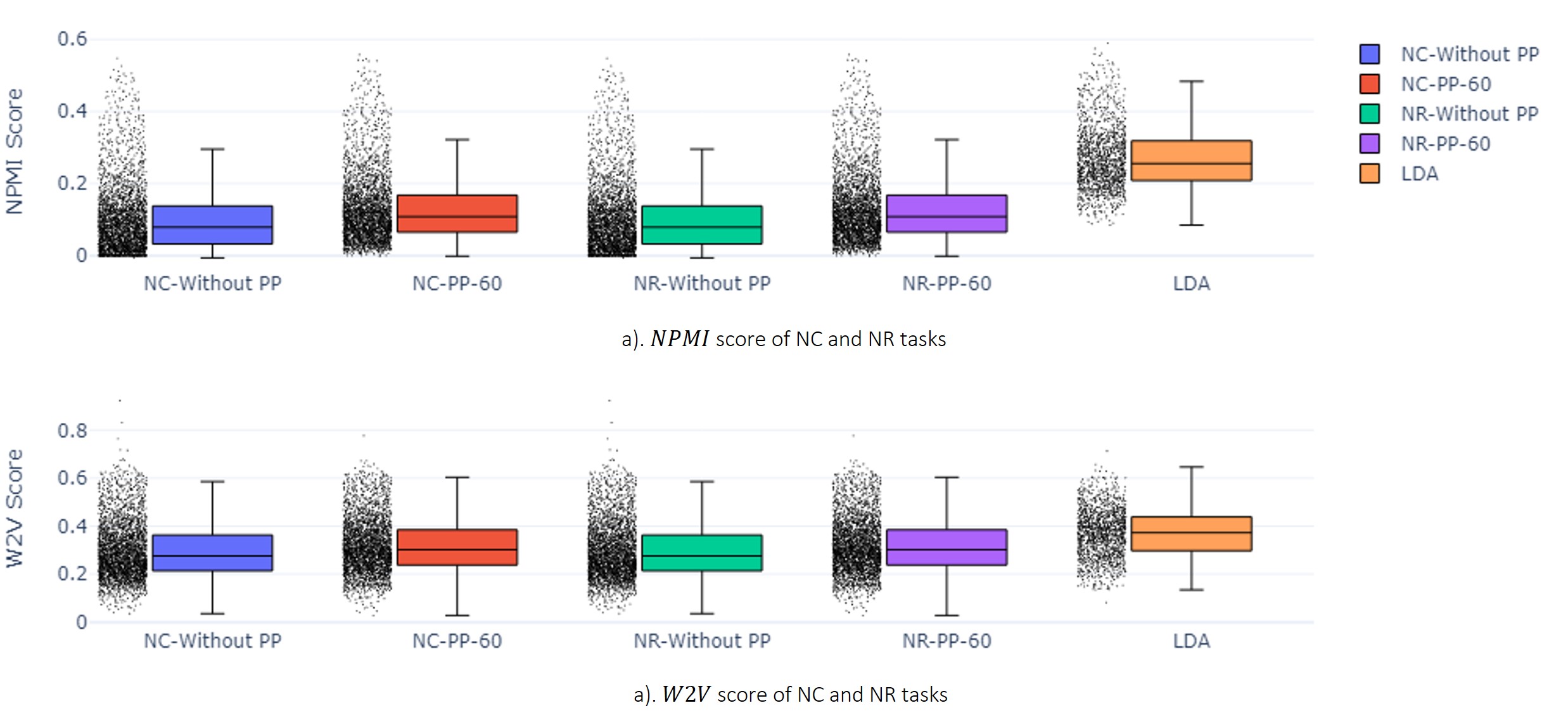}
  \caption{Box plot for the distribution of $NPMI$ scores and $W2V$ similarity scores for our $BATM\text{-}ATT$ model on the news classification (NC) and news recommendation (NR) tasks.}
  \label{fig:box_plots_nc_nr}
  \Description{A box plot}
\end{figure}

\begin{figure}[!t]
  \centering
  \includegraphics[width=\linewidth]{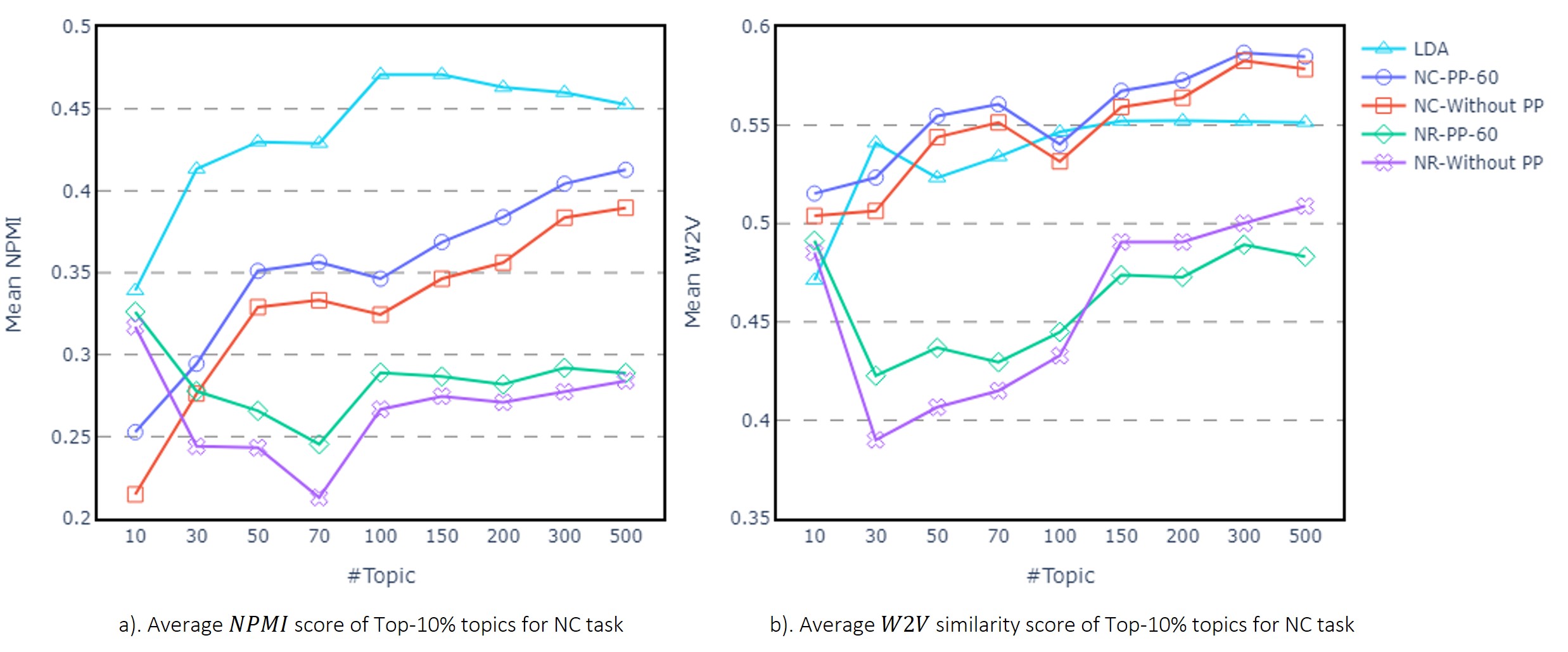}
  \caption{Plots of average top-10\% topics' $NPMI$ scores and $W2V$ scores for increasing numbers of topics. We consider both the news classification (NC) and news recommendation (NR) tasks. We also include scores for a standard  LDA topic model for comparison.}
  \label{fig:top10_averages_nc_nr}
  \Description{A box plot}
\end{figure}

In Fig.~\ref{fig:box_plots_nc_nr} we see the distributions of $NPMI$ scores and $W2V$ similarity scores computed from our $BATM\text{-}ATT$ model, where each point in a plot represents a one-time evaluation of topic descriptors. Here the number of topics ranges between $10\text{-}500$. We can make several observations from the box plots. Firstly, post-processing slightly improves scores and decreases the number of poor topics. This is because our model is not highly dependent on the processing pipeline and can robustly deal with non-informative words like stopwords that usually will not appear in the topic descriptors generated by our models. However, the classical LDA model relies on strict preprocessing procedures to avoid including non-informative words in descriptors. Secondly, the LDA model is significantly better than our $BATM\text{-}ATT$ model in extracting coherent topics in terms of the $NPMI$ metric, but our model achieves comparable results on the $W2V$ metric. This is reasonable because our aim here is to generate explanations, rather than focusing on constructing complete topic models for corpus exploration. Finally, our model can often generate coherent topics comparable to the LDA model, where more than 10\% of the topics have a $NPMI$ score above 0.2 and a $W2V$ similarity score over 0.4 as well. This inspires us to compare average scores of only the top-10\% topics as we normally only care about several most important topics and generate explanations based on these topics. According to Fig.~\ref{fig:top10_averages_nc_nr}, the trend of top-10\% average NPMI scores and $W2V$ similarity scores is $LDA > NC > NR$. Since the complexity of the NR task requires user modeling based on the news modeling, which may be negative in finding meaningful topics because of the involvement of unsure factors, the extracted topics are not as good as the LDA model. However, those topics are enough to explain recommendation behaviors as we only require some of them in a real case. These results validate the effectiveness of our BATM\text{-}ATT model in generating high-quality topics, 

\subsection{Case Study}
\label{sec:case_study}
\begin{figure}[!ht]
    \centering
    \includegraphics[width=\linewidth]{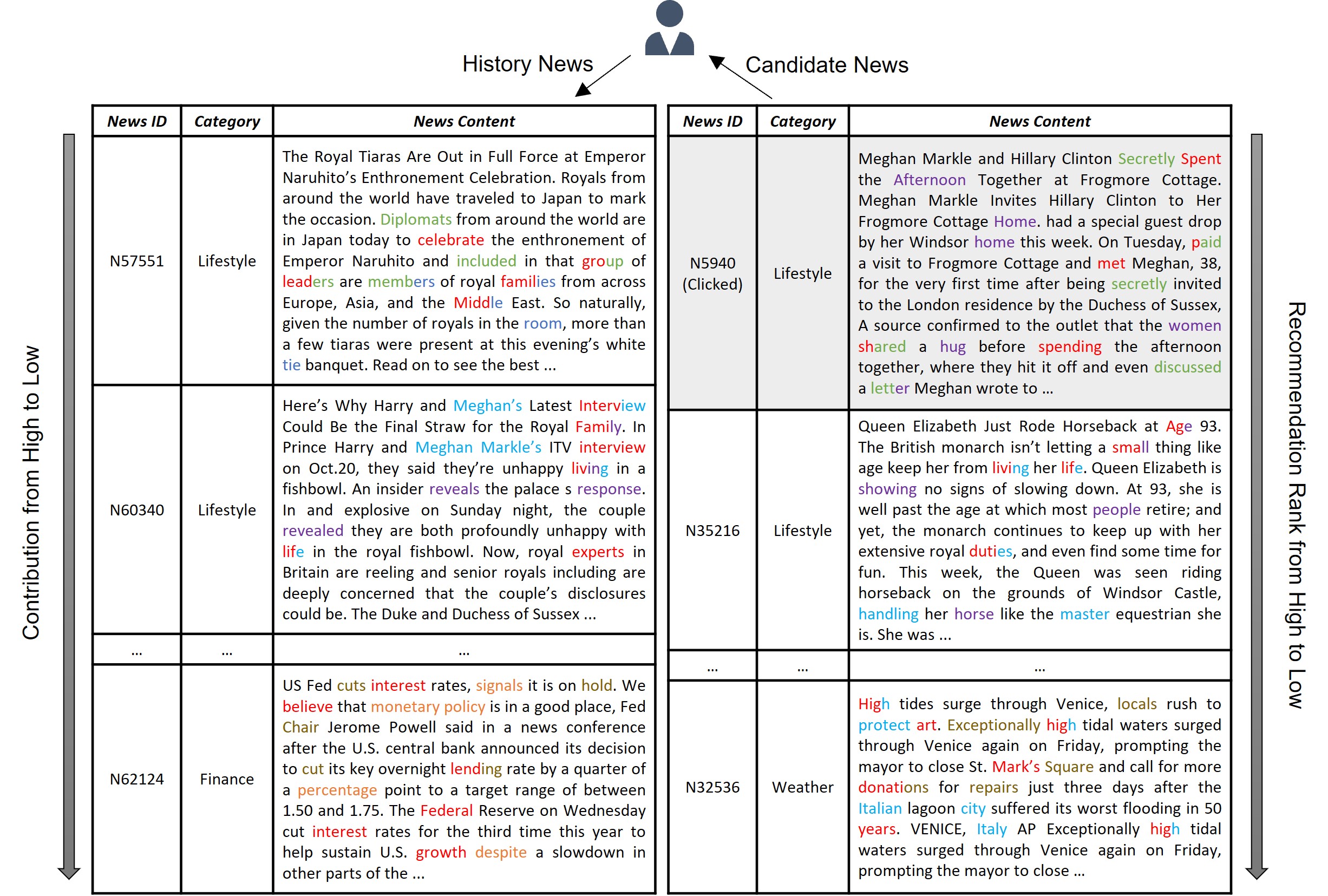}
    \caption{A typical example of a recommendation case selected from the MIND test set. The topics for the news articles are highlighted with a consistent set of colors, each corresponding to a different topic. Since a word can potentially be associated with more than one topic, some words may have two or three colors. A color that appears more often signifies that the corresponding topic contributes more to an article than the other topics.}
    \label{fig:case_study_good}
\end{figure}

We conduct a case study to present how our BATM-ATT model generates explanations of a real-world example using the method of Section~\ref{sec:rec_explain} as shown in Fig.~\ref{fig:case_study_good}. A case involves the user browsed articles $\mathcal{H}$ and the ranked candidate news articles $\mathcal{C}$ shown to the user. We highlight those descriptor words from three of the most important topics for each news article, where each topic is highlighted with a unique color. 

Firstly, we notice that the topic in red $T_{red}$ is activated across nearly all articles. The exact meaning of the topic $T_{red}$ is hard to summarize, as it is always highly related to the specific article itself. For example, it focuses on words such as "interview", "family", "living", and "life" for the article \textit{N60340}, which are about daily life. However, for a financial news article \textit{N62124}, the highlighted words now include "interest", "believe", "lending", and "federal", which are often used in the financial domain. We also observe that news articles under the same category usually discuss similar topics, and we observed that the example case also reflects this nature. Take the category "lifestyle" as an example, except for the general topic $T_{red}$, there are four more topics that concern words related to lifestyle for all four news articles. Among these four topics, the topic in green $T_{green}$ and the topic in purple $T_{purple}$ only appear in the articles on lifestyle. Thus, the last observation is that some topics are closely related to a specific subject, such as the topic in orange $T_{orange}$, which is highly related to the political economy topic, and attends to words like "signals", "monetary" and "policy". For further examples of topics generated by our approach, see the Appendix.

\section{Conclusion and Future Work}

In this paper, we presented a novel recommender architecture that harnesses a bi-level attention framework to decouple the news recommendations process as topic capturing, topic importance recognition, and decision-making process to benefit explainability. We conducted experiments on a real-world news recommendation dataset where we compared our approach to  several state-of-the-art alternatives. Results indicate that our model can achieve better performance while also successfully capturing intuitive meanings in the form of topical features, thus improving its explainability and transparency. Furthermore, we applied two topic coherence metrics to quantitatively evaluate the quality of the topics generated by our model, in the context of both news classification and recommendation tasks, thereby validating the interpretability of our model. For future work, we suggest three distinct directions. First, we intend to explore the use of topic features to improve news recommendation performance. Second, we plan to add constraints to our model to extract more interpretable topics. Finally, we intend to conduct a user study to determine how well our explainable recommender works in a real-world setting.

\section{Appendices}

\subsection{Experimental Environment}
Our experiments were conducted on a High Performance Computing cluster running the Linux operating system. We used PyTorch 1.8.0 as the backend. The GPU type is Nvidia Tesla V100 and A100 with 32GB and 40GB GPU memory respectively. We ran each experiment 5 times with fixed random seeds, each in a single thread.

\subsection{Additional Case Study}

\begin{figure}[!ht]
    \centering
    \includegraphics[width=\linewidth]{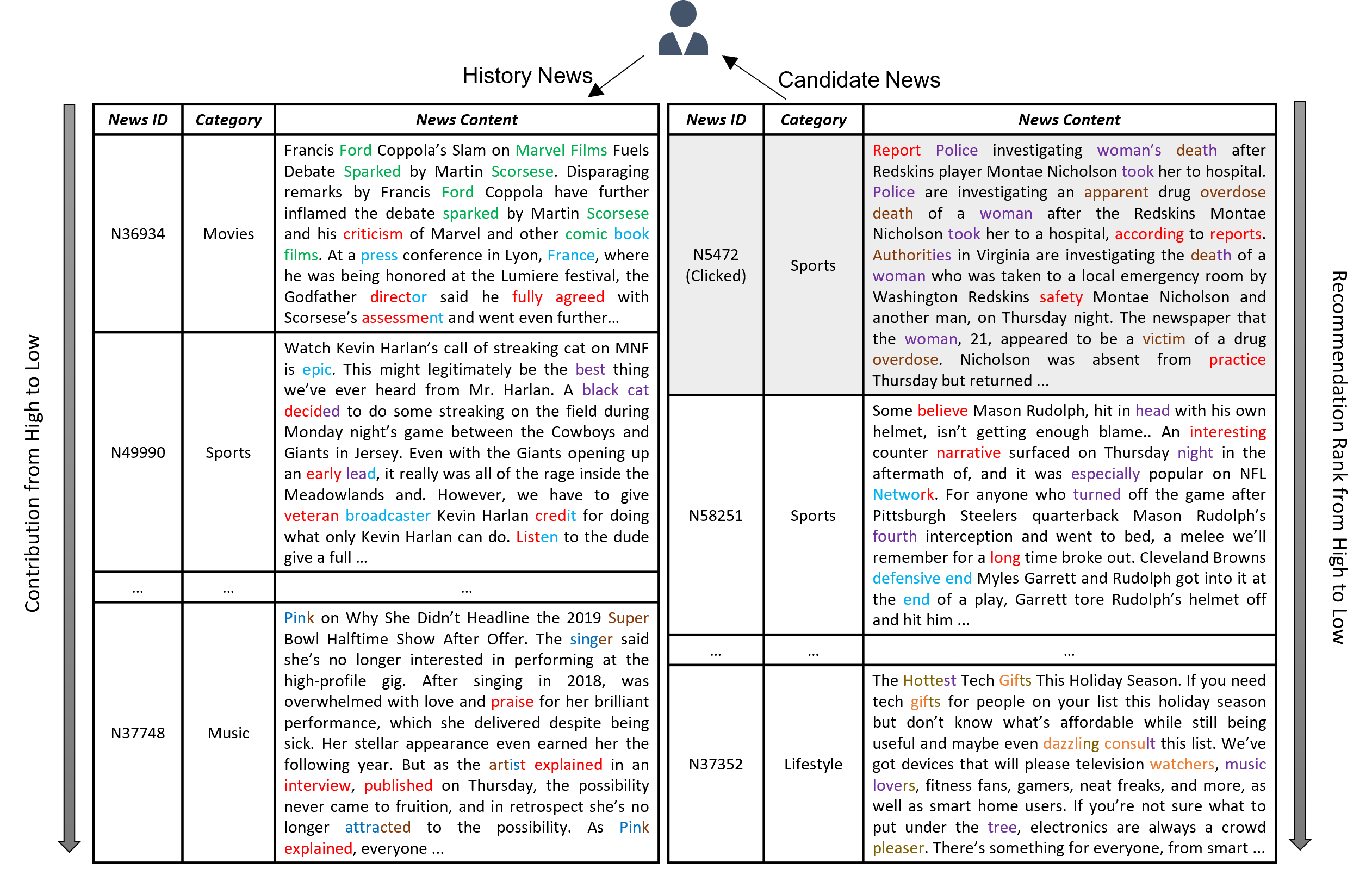}
    \caption{Another case with the same format as Fig.~\ref{fig:case_study_good}, sampled from the MIND test set.}
    \label{fig:case_study_incoherent}
\end{figure}

In Fig.~\ref{fig:case_study_incoherent} below we show another typical example of a recommendation case sampled from the MIND test set.

\subsection{Global Topic Examples}
\begin{table*}[!ht]
    \centering
    \begin{tabularx}{\linewidth}{X l l}
        \toprule
         \thead{Topic Descriptor} & \thead{$C_{NPMI}$} & \thead{$C_{W2V}$} \\
        \midrule
        dog cat terrier canine kennel pup retriever dogs pups canines &  \multirow{1}{*}{0.5131}&  \multirow{1}{*}{0.6544} \\
        \addlinespace[3pt]
        song songs album music guitar piano sound soundtrack audio nice &  \multirow{1}{*}{0.3526}&  \multirow{1}{*}{0.5055} \\
        \addlinespace[3pt]
        undergraduate admissions faculty universities graduate bachelor university students enrolled colleges &  \multirow{2}{*}{0.3546}&  \multirow{2}{*}{0.6040} \\
        \addlinespace[3pt]
        loan loans mortgages million mortgage river download borrowers lenders equity &  \multirow{1}{*}{0.3201}&  \multirow{1}{*}{0.4386} \\
        \addlinespace[3pt]
        pastas entrees salmon seafood appetizer mussels appetizers shrimp lobster oysters &  \multirow{1}{*}{0.5793}&  \multirow{1}{*}{0.5952} \\
        \addlinespace[3pt]
        oven ingredients bake recipes cooking baking protein crust butter recipe &  \multirow{1}{*}{0.5937}&  \multirow{1}{*}{0.5214} \\
        \addlinespace[3pt]
        \bottomrule
    \end{tabularx}
    \caption{Examples of topics identified by our approach, in terms of extracted topic descriptors, with coherence scores calculated via $C_{NPMI}$ and $C_{W2V}$.}
    \label{tab:appendix_topic}
\end{table*}

\begin{acks}
This research was supported by Science Foundation Ireland (SFI) under Grant Number SFI/12/RC/2289\_P2. 
\end{acks}

\bibliographystyle{ACM-Reference-Format}
\bibliography{reference}

\appendix

\end{document}